\documentclass[9pt,twocolumn,twoside]{opticajnl}
\journal{opticajournal} % use for journal or Optica Open submissions

\setboolean{shortarticle}{false}
\title{Single-shot pulse retrieval of femtosecond bright squeezed vacuum}

\doi{}

\author[1,2]{Yuval Kern}
\author[1,2]{Ido Nisim}
\author[3,2]{Michael Birk}
\author[4,5]{Andrei Rasputnyi}
\author[1,2]{Doron Behar}
\author[1,2]{Zhaopin Chen}
\author[6,2]{Ido Kaminer}
\author[6,2]{Pavel Sidorenko}
\author[1,2,7]{Oren Cohen}
\author[1,2,*]{Michael Krüger}

\affil[1]{Department of Physics, Technion -- Israel Institute of Technology, 32000 Haifa, Israel}
\affil[2]{Solid State Institute and Helen Diller Quantum Center, Technion -- Israel Institute of Technology, 32000 Haifa, Israel}
\affil[3]{The Russell Berrie Nanotechnology Institute, Technion -- Israel Institute of Technology, 32000 Haifa, Israel}
\affil[4]{Max Planck Institute for the Science of Light, 91058 Erlangen, Germany}
\affil[5]{Friedrich–Alexander Universität Erlangen–Nürnberg, 91058 Erlangen, Germany}
\affil[6]{Electrical and Computer Engineering, Technion -- Israel Institute of Technology, 32000 Haifa, Israel}
\affil[7]{Department of Physics, Guangdong Technion -- Israel Institute of Technology, Shantou 515063, Guangdong, China}

\affil[*]{krueger@technion.ac.il}

\begin{abstract}
Bright squeezed vacuum (BSV) is an intense quantum state of light with zero mean electric field and huge photon number fluctuations, sufficiently intense to drive extreme nonlinear processes and imprint nonclassical statistics. However, the temporal structure of single BSV shots has not been fully characterized. Here, we retrieve the spectral and temporal pulse characteristics of a set of single-peak BSV shots. It is obtained by realizing a femtosecond BSV source at 1040\,nm with a single spatial mode and perform single-shot spectral interferometry with a fully characterized coherent-state reference pulse. Our approach reveals that the group delay is consistent between the various shots, resulting in an average pulse duration of $27.2$\,fs, much shorter than the pump pulse, and a variation of 5.5\,fs (standard deviation). We also observe a characteristic nodal structure in the spectral interferograms, demonstrating the BSV’s random phase ambiguity of $\pi$\,rad. Our approach demonstrates that BSV is a viable source of femtosecond light pulses for attosecond sub-cycle metrology of ultrafast electron dynamics.
\end{abstract}

\setboolean{displaycopyright}{false} % Do not include copyright or licensing information in submission.

\begin{document}

\maketitle

\section{Introduction}

\begin{figure}[htb]
    \centering
    \includegraphics[width=\columnwidth]{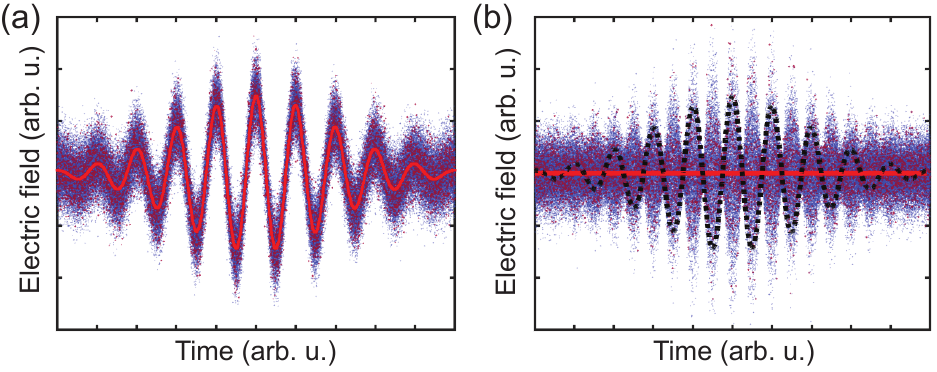}
    \caption{Illustration of bright squeezed vacuum (BSV). (a) Illustration of the electric field and its fluctuations for a ``classical'' (coherent-state) state laser pulse. Red curve: Average electric field. (b) The same for BSV. While any single realization of BSV field (dotted black curve) has a defined waveform, the field fluctuations across many possible pulses can be positive or negative with equal probability, resulting in an average electric field of zero (red line). }
    \label{fig1}
\end{figure}

Ultrafast science in the attosecond temporal regime is based on extremely nonlinear sub-cycle processes driven by intense laser pulses, with each driving laser pulse containing $10^{7}$-$10^{15}$ photons~\cite{Ferray1988,Corkum1993,Lewenstein1994,Hentschel2001,Paul2001,Corkum2007,Kruger2011,Ghimire2011,Vampa2015a,Heide2024}. This is in stark contrast to quantum optics with nonclassical light, which is commonly associated with photon statistics of up to $10^2$ photons~\cite{Boyd2019,Erhard2020}. Introducing concepts of quantum optics in ultrafast science, such as squeezing~\cite{Teich1989}, needs to bridge this stark intensity gap. Recently, the development of intense sources of nonclassical light, particularly pulses of bright squeezed vacuum (BSV)~\cite{Iskhakov2012,Spasibko2017,Manceau2019}, has demonstrated that this gap can be overcome. BSV pulses possesses fascinating quantum properties, which distinguishes it strongly from ``classical'' (coherent-state) laser pulses used in ultrafast science. In the time domain, ideal coherent-state classical laser pulses exhibit oscillations of the electric field with the fundamental frequency with small fluctuations due to the quantum vacuum, characterized by Poissonian photon statistics. In BSV,  however, the average electric field is zero, in line with its character as a vacuum state, whereas the electric field fluctuations are extremely large and oscillate at twice the light frequency, in line with the squeezed nature of BSV (see Figure~\ref{fig1}(a) and (b) for an illustration). The BSV photon statistics support a plethora of different realizations of pulse shapes. Individual BSV shots can reach extremely high photon numbers of up to $10^{13}$ within a single BSV pulse due to  antisqueezing~\cite{Spasibko2017,Manceau2019,Rasputnyi2024}. Notable is also a phase ambiguity of $\pi$\,rad and the nodal structure of the statistics at the field zero crossings, which follows from the oscillations of the fluctuations at twice the BSV light frequency. Average pulse durations down to 25\,fs in BSV at 1600\,nm have recently been reported~\cite{Rasputnyi2024}.

Femtosecond BSV pulses have been shown to drive nonlinear processes with higher efficiency than coherent light with the same mean intensity, due to the fact that the nonlinearity of those processes enhances the statistical contribution of the BSV shots with high photon numbers, though those shots are rare. Perturbative nonlinear effects, such as low-order harmonic generation~\cite{Spasibko2017} and multiphoton electron emission from nanotips~\cite{Heimerl2024} have been driven solely by BSV, resulting in increased efficiency. BSV has also been applied as a perturbation field for high-harmonic generation (HHG) from solids~\cite{Lemieux2025} and gases~\cite{Tzur2025} as a route to imprint non-classical photon statistics on the resulting harmonics in the UV and extreme UV spectral regions. Nonperturbative HHG from solids~\cite{Rasputnyi2024} and strong-field photoemission from nanotips~\cite{Heimerl2025} solely driven by BSV open up interesting effects in the sub-cycle strong-field regime of light-matter interactions, as predicted by theory~\cite{Gorlach2023}. The use of BSV to introduce nonclassicality is part of a larger experimental and theoretical effort to bridge ultrafast science and quantum optics through different pathways, such as electro-optic approaches~\cite{Riek2015,Riek2017,Virally2021,BeneaChelmus2025}, HHG interactions~\cite{Gorlach2020,Lewenstein2021,Tzur2023,Tzur2024,Stammer2024,Tzur2025} and photoionization experiments~\cite{Koll2022,Laurell2025}.

While BSV is well-established and can be characterized using quantum optical methods (see, e.g.,~\cite{Iskhakov2012,Iskhakov2012a,Spasibko2017,Manceau2019}), sub-cycle ultrafast studies require a complete characterization of the electric field in time for each shot of femtosecond BSV. Each shot is a different realization of the statistical quantum properties of BSV, i.e., it corresponds to a femtosecond light pulse with well-defined spectral intensity and phase. Although the spectral intensity can be easily measured with a spectrometer with single-shot capability, the spectral phase is challenging to measure, particularly in a single-shot measurement together with the intensity. In~\cite{Rasputnyi2024}, frequency-resolved optical gating (FROG) was applied to femtosecond BSV pulses at 1600\,nm wavelength but could only reveal an average pulse shape with a duration of about 25\,fs. 
%While FROG and also spectral phase interferometry for direct electric-field reconstruction (SPIDER) can be extended to single-shot acquisition and pulse retrieval, the strongly fluctuating shot-to-shot intensity variation of BSV demands a large dynamical range of the measurements which is challenging to achieve with these nonlinear methods.
Here, we introduce a conceptually simple yet powerful approach to measure the spectral phase of single shots of BSV -- single-shot spectral interferometry. We implement a femtosecond BSV source at a wavelength of 1040\,nm with a single spatial mode and interfere it with a broadband synchronized classical laser pulse that has been fully characterized. We analyze the spectral interference fringes and retrieve the spectral phase of each BSV shot, which enables us, for the first time, to characterize BSV in pulse duration and its shot-to-shot variations. Our measurements reveal that the average pulse duration of the shots corresponding to the fundamental spectral mode of the BSV is around 27.2\,fs. We choose to use spectral interferometry because it is simple and has the advantage that it does not require a numerical iterative algorithm and can handle low light intensities in contrast to nonlinear approaches to measure pulses.

\section{Results}

\subsection{Bright-squeezed-vacuum source at 1040 nm}

In this section, we introduce our BSV source with a central wavelength of 1040\,nm. This is a shorter wavelength than 1600\,nm that has been used for previous ultrafast light-matter-interaction BSV experiments~\cite{Rasputnyi2024,Lemieux2025,Tzur2025,Heimerl2025} and aligns with the ubiquitous use of Yb-based laser systems in ultrafast science~\cite{Truong2025}. The experimental approach to produce BSV is to strongly pump an unseeded optical parametric amplifier (OPA) phase-matched for a collinear frequency-degenerate mode. Due to the absence of seed light, fluctuations of the quantum vacuum are amplified by the OPA, starting with photon-pair generation leading to the emission of extremely bright light pulses. In our experiment, we employ the 515-nm second harmonic (SH) of 1030\,nm, 178\,fs pulses from an Yb-based amplified laser system as a pump (see Figure~\ref{fig1}(b) for a sketch of the setup). The collimated SH pump pulses ($\sim$\,125\,fs pulse duration, $121\,{\mu}$J pulse energy, 1\,kHz repetition rate) is sent into a 1 mm thick $\beta$-barium borate (BBO) crystal for Type I phase-matching cut at $\theta = 23.4^\circ$. With no seed present, a weak BSV beam is generated. Since the BSV beam possesses many spatial modes, a filtering approach needs to be applied. Here we apply the second stage of amplification by placing the second identical BBO crystal in the beam path at a distance of 22.8\,cm~(cf.~\cite{Perez2014}). This distance ensures that the higher-order spatial modes of the BSV are not amplified in the second BBO crystal. Only the fundamental Gaussian spatial mode overlaps well with the collimated 515-nm pump beam and is amplified. The distance also guarantees that the relative phase between 515-nm light and BSV is such that the amplification in the second BBO crystal is constructive. We obtain spatially single-mode BSV light at a central wavelength of 1040\,nm. This wavelength is consistently obtained when optimizing the power of the BSV through tuning the phase-matching angles of both crystals. We observe a nonlinear frequency shift of the 515-nm light to longer wavelength, consistent with the observation of 1040\,nm BSV light (see Supplementary Material). The BSV possesses extreme intensity fluctuations and an average pulse energy of $\sim$\,100\,nJ, corresponding to $5 \times 10^{11}$ photons on average. A gain curve measurement shows the absence of saturation of the parametric amplification (see Supplementary Material).

\begin{figure}[!htb]
    \centering
    \includegraphics[width=0.95\columnwidth]{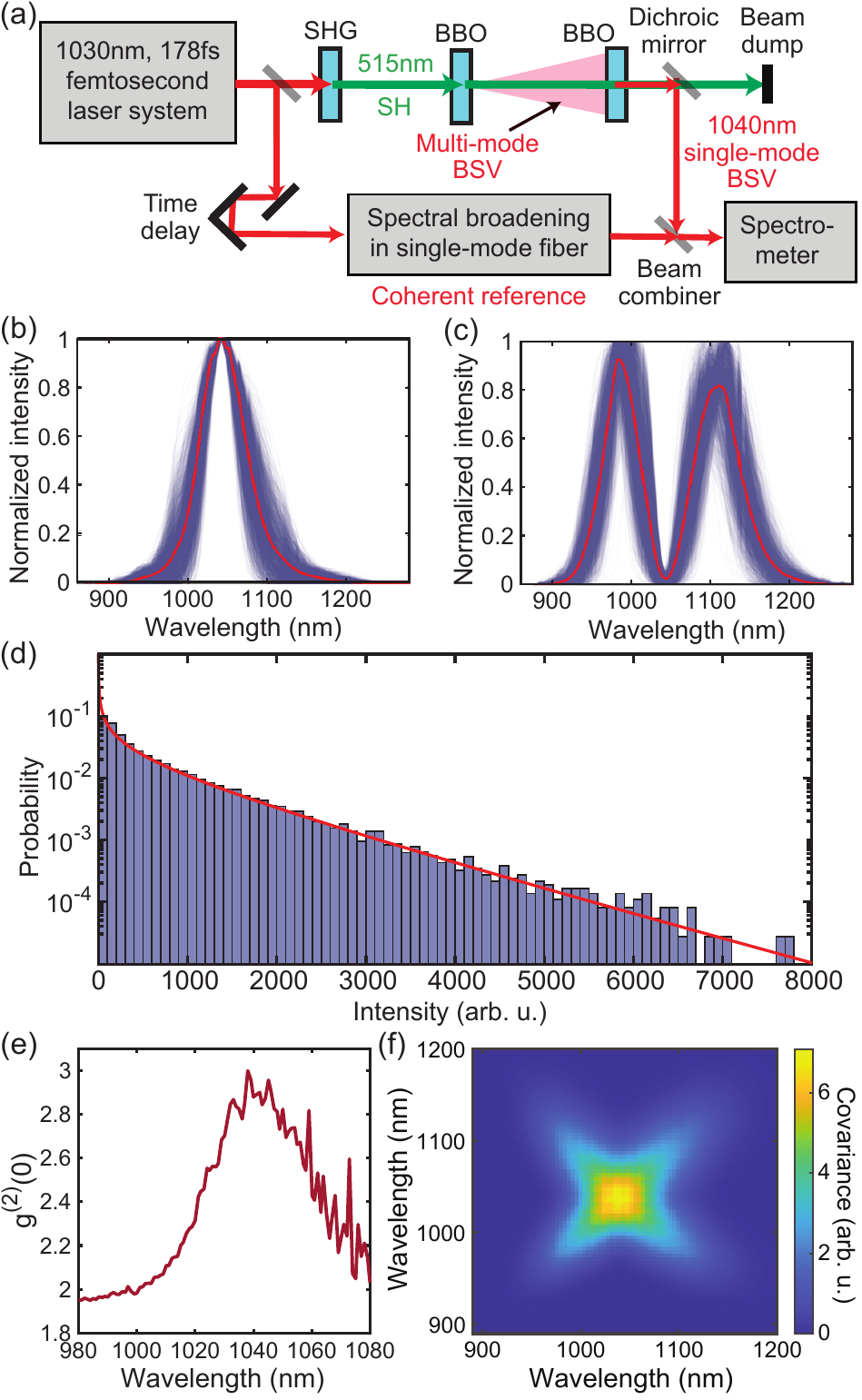}
    \caption{Generation and characteristics of 1040-nm BSV. (a) Experimental setup. SH and SHG: second harmonic and second harmonic generation, respectively; BBO: $\beta$-barium borate crystal. (b) Normalized spectra of BSV shots with the fundamental spectral mode (blue) and their average (red). (c) The same, but for the double-peak spectral mode. (d) Intensity statistics at 1040\,nm (red curve: Fit with BSV photon statistics). (e) Second-order correlation parameter $g^{(2)}(0)$ of the generated BSV light as a function of wavelength. (f) Spectral covariance plot.}
    \label{fig2}
\end{figure}

We measure the spectrum of each individual femtosecond BSV pulse using a near-infrared InGaAs-based spectrometer. Normalizing the spectra, we find three characteristic spectral shapes, a single peak centered at $\sim$\,$1040$\,nm (Figure~\ref{fig2}(b)), a two-peak structure with a minimum at $\sim$\,$1040$\,nm (Figure~\ref{fig2}(c)) and mixtures of the two (not shown). Single-peak spectra indicate the fundamental spectral mode of the BSV with a roughly Gaussian shape, whereas the double-peak structures correspond to the next higher spectral mode. The double-peak structure indicates the generation of a higher-order Hermite-Gaussian frequency mode.

In the next step, we characterize the intensity statistics of our 1040-nm BSV pulses. We record 40,000 single-shot BSV spectra using a laser-triggered Si-based spectrometer and determine $g^{(2)}(0)$, normalized second-order correlation function at zero time delay, as a function of wavelength $\lambda$. For large intensities, it is given by $g^{(2)}(0) \approx \left<N^2(\lambda)\right> / \left<N(\lambda)\right>^2$ where $N(\lambda)$ is the (shot-resolved) spectral intensity at a given $\lambda$ and $\left< \ldots \right>$ indicates the statistical mean. Figure~\ref{fig2}(d) shows that $g^{(2)}(0)$ approaches 3 at $\sim$\,$1040$\,nm in the center of the BSV spectrum as expected for degenerate squeezed vacuum~\cite{Breitenbach1997}. Further away from the center, $g^{(2)}(0)$ decreases to $\sim$$2$ which is consistent with a thermal light state resulting from the fact that the nondegenerate double-peak spectral mode starts to contribute.

We shed more light on the properties of our BSV source by determining its spectral intensity correlations. We employ the InGaAs-based spectrometer and record $4 \times 10^4$ single-shot spectra. Figure~\ref{fig2}(f) shows the resulting spectral intensity covariance  $\text{Cov}(\lambda_1, \lambda_2) = \left<N(\lambda_1)N(\lambda_2)\right> - \left<N(\lambda_1)\right>\left<N(\lambda_2)\right>$. In addition to the trivial auto-correlation on the rising diagonal, we also find a cross-correlation and a strong central peak where both overlap at $\sim$\,$1040$\,nm. The orthogonal component nicely visualizes the correlation of signal and idler photons in our vacuum-seeded OPA. The central peak indicates the spectral range in which photons can be considered indistinguishable, in agreement with the $g^{(2)}(0)$ measurement in Fig.~\ref{fig2}(e). A comparison of the conditional spectral width with the unconditional spectral width reveals the presence of 1.62 spectral modes.

\subsection{Single-shot spectral interferometry}

\begin{figure}[!htb]
    \centering
    \includegraphics[width=\columnwidth]{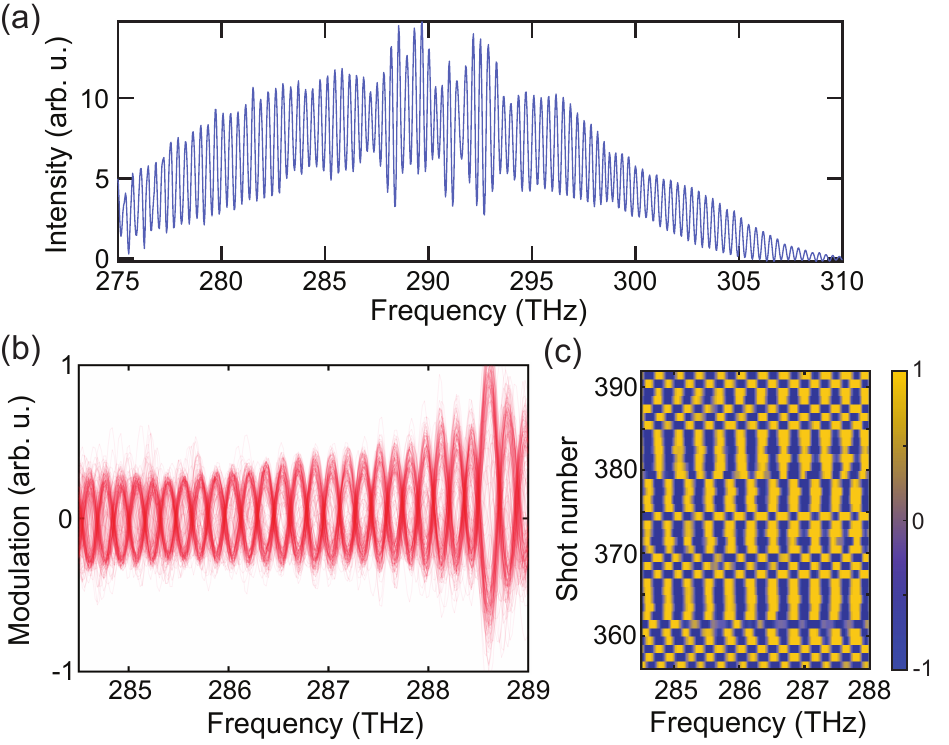}
    \caption{Single-shot spectral interferometry with BSV. (a) Typical single-shot spectral interferogram at a delay of 3.05\,ps. (b) Modulation spectrum displaying the interference fringes and the BSV's nodal structure plotted for a sequence of 200 shots. We normalized the fringes and subtracted the DC offset at a frequency of 286\,THz. (c) Color plot of the modulation spectrum as a function of shot number for a sequence of 35 shots.}
    \label{fig3}
\end{figure}

Here we focus on the ultrashort spectro-temporal properties of each individual BSV shot. Our measurement focuses solely on classical properties, like the variance of the anti-squeezed quadrature, while the variance of the squeezed quadrature can not be resolved due to the lack of the dynamical range of spectrometer. In order to be able to obtain the spectral phase and amplitude for each shot, we implement spectral interferometry between the unknown BSV pulses and a stable, fully characterized coherent-state light pulse as a reference (see Supplementary Material). To this end, we split off a part of the fundamental laser light at 1030\,nm and spectrally broaden it through self-phase modulation (SPM) in 4.2\,cm of a standard single-mode fiber (PM980)~\cite{Stolen1978,Liu2016}. The spectral broadening stage is necessary because we need to cover the bandwidth of the BSV spectra as much as possible. At a pulse energy of 45\,nJ, the resulting reference pulse covers 960-1100\,nm which is slightly narrower than the bandwidth of the single-peak spectral mode. We recorded the reference pulse with a commercial frequency-resolved optical gating (FROG) device and determined its spectral phase (see Supplementary Material). We interfere the reference pulse with the BSV using a 20:80 beam recombiner and record their interferogram with the Si-based spectrometer with single-shot capability (see Fig.~\ref{fig2}(a) for an illustration). A temporal delay of $\sim$3.05\,ps leads to a dense spectral interference fringe pattern (see Fig.~\ref{fig3}(b) for an example).

The single-shot spectral interference fringes directly reveal the $\pi$ phase ambiguity, an important property of the BSV. The quantum vacuum that we amplify through the unseeded OPA process initially contains all phases. However, according to the classical and quantum understanding of the phase-sensitive OPA, the amplification process in degeneracy can occur both in 0 and $\pi$ phase difference between the pump and the down-converted light~\cite{Yariv1991}. This leads to the $\pi$ phase ambiguity, which manifests itself directly in the fringe pattern recorded in our experiment. Figure~\ref{fig3}(b) shows the fringe pattern after subtraction and normalization for a sequence of 200 pulses. We observe a pronounced nodal structure due to the fact that the zero crossing occurs at the same frequencies for each shot. The relative distribution of the two phases is 105:95, corresponding to a binary random distribution with probability $0.525\pm0.035$, which is compatible with randomness in the phase of the quantum vacuum.  Our measurement is limited here by the fact that the interferometric delay in our setup is not stabilized. A different visualization of this effect is depicted in Fig.~\ref{fig3}(c) where a checkerboard pattern emerges in some instances.

\subsection{Single-shot BSV pulse reconstruction}

Our measurement of single-shot spectral interferograms allows for a full reconstruction of the spectro-temporal properties of each BSV shot. Here we focus on 1009 single-peak spectra out of 16,000 recorded shots due to their excellent spectral overlap with the reference pulse. Criteria for selecting single-peak spectra include the number of peaks, peak wavelength and spectral width (for details see the Supplementary Material). Calculating the Fourier transform of the interferograms allows us to isolate DC and AC terms, which give us access to the spectrum of each BSV shot and its relative phase with respect to the reference pulse, respectively~\cite{Takeda1982} (see Supplementary Material). 

Figure~\ref{fig4}(a) shows the results of our reconstruction in the spectral domain. We find that the normalized single-peak spectra differ in both peak position and spectral width. More important is the reconstructed group delay of the BSV shots, also plotted in Fig.~\ref{fig4}(a). On top of the average group delay we find little variation between the individual shots near the central frequency of 288\,THz, with increasing noise at the edges of our measurement range. The average group delay exhibits fine oscillations and peaks which we attribute to imperfections in the FROG reconstruction of the reference pulse. We also note that the interferograms are recorded at the spectrometer, which means that the dispersion effects of optical elements on the beam path between BSV generation and detection are included in the measurement. 

\begin{figure}[!htb]
    \centering
    \includegraphics[width=0.9\columnwidth]{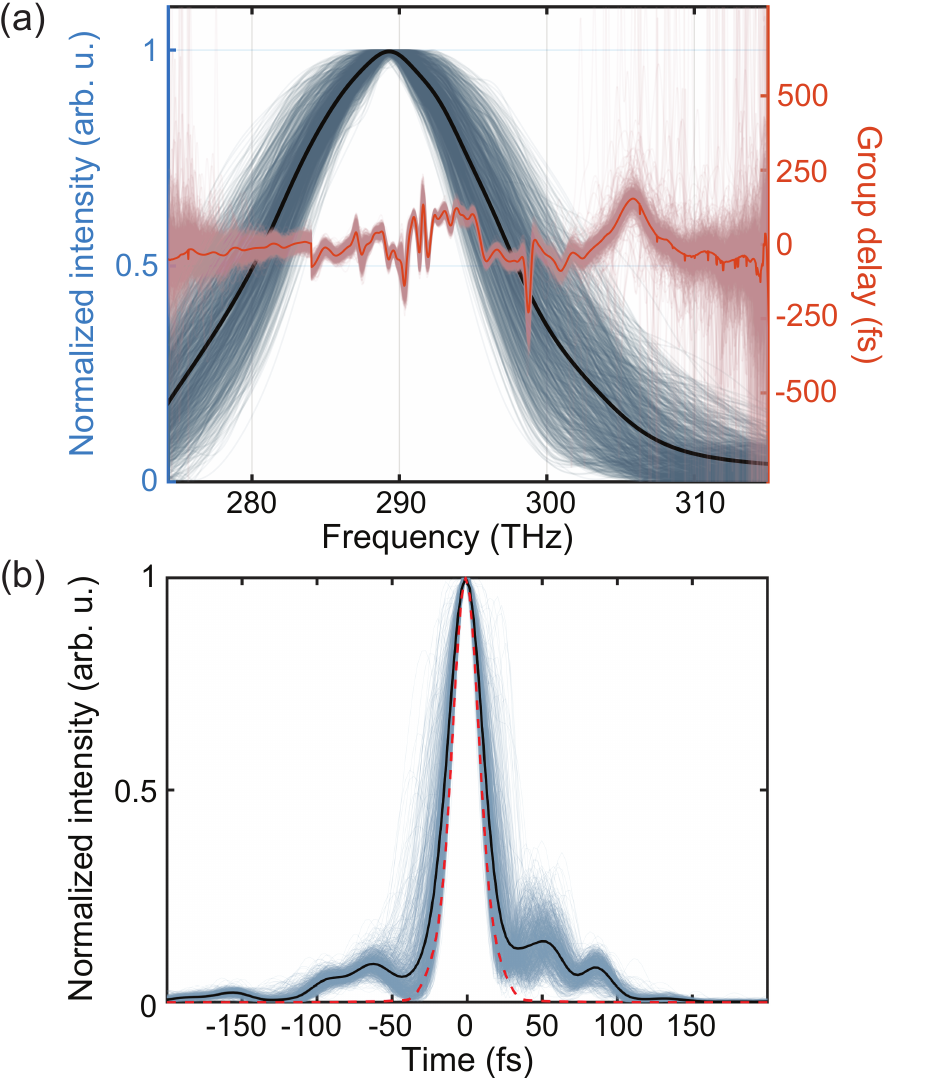}
    \caption{Reconstruction of single-peak BSV shots. (a) Retrieved normalized spectral intensity (blue-grey) and group delay (light orange) as a function of frequency. The black curve and the orange curve represent the average of the spectral intensity and group delay, respectively. (b) Time-domain reconstruction of the intensity profile of the shots (blue-grey). The black curve and dashed orange curve are the average pulse profile and the transform-limited pulse profile calculated from the average spectrum from (a).}
    \label{fig4}
\end{figure}

In the time domain, we find an average pulse duration of 27.2\,fs (full width at half maximum intensity), while the transform-limited pulse duration corresponding to the average spectrum is 19.3\,fs. Remarkably, this is much shorter than the pulse duration of 178\,fs delivered by the laser system at its output or the estimated pulse duration of 126\,fs of its second harmonic, which serves as the pump. The exponential scaling of BSV with pump intensity concentrates the emission in time, causing it to occur predominantly at the peak of the field's pulse envelope. The standard deviation of the pulse duration of the analyzed set of BSV shots is 5.5\,fs, reflecting the variations in spectral width and group delay. The side peaks are likely due to to imperfections in the FROG reconstruction of the reference pulse.

\section{Conclusion}

In conclusion, we have demonstrated a single-shot resolved retrieval of the spectro-temporal profile of a femtosecond BSV source at 1040\,nm. We obtain an average pulse duration of 27.2\,fs, much shorter than pump pulse. We also observe a $\pi$\,rad phase ambiguity, in line with the classical properties of BSV. The capability to retrieve the temporal structure of the electric field is an important prerequisite for attosecond science experiments. BSV has been shown to enable measurements of electron dynamics beyond the conventional damage threshold~\cite{Rasputnyi2024}. Here we envision that BSV will be at the heart of a sub-cycle spectroscopic scheme where it interacts strongly with matter and we retrieve the shot-resolved waveform after the interactions, detecting the temporal imprints of nonlinear polarization~\cite{Sommer2016}, photoionization and dielectric breakdown~\cite{Husakou2024}, and phase transitions~\cite{Valmispild2024}.

\begin{backmatter}

\bmsection{Funding} This work has received funding from the Israel Science Foundation under grant no.~1315/24. We acknowledge partial financial support through the Quantum Flagship program of the Helen Diller Quantum Center at the Technion and through the collaborative quantum research program at University of Toronto and the Technion.

\bmsection{Acknowledgment} We thank Maria V.~Chekhova and Barak Dayan for insightful discussions.

\bmsection{Disclosures} The authors declare no conflicts of interest.

\bmsection{Data availability} Data underlying the results presented in this paper are not publicly available at this time but may be obtained from the authors upon reasonable request.

\bmsection{Supplemental document}
See Supplement 1 for supporting content.
\end{backmatter}

%\section{References}

%Note that \emph{Optics Letters} and \emph{Optica} short articles use an abbreviated reference style. Citations to journal articles should omit the article title and final page number; this abbreviated reference style is produced automatically when the \emph{Optics Letters} journal option is selected in the template, if you are using a .bib file for your references.

%\bigskip
%\noindent Add citations manually or use BibTeX. See \cite{Zhang:14,OPTICA,FORSTER2007,testthesis,manga_rao_single_2007}. List up to three author names in references. If there are more than three authors, list the first three followed by \emph{et al.}

% Bibliography
%\bibliography{scibib}

% Full bibliography added automatically for Optics Letters submissions; the following line will simply be ignored if submitting to other journals.
% Note that this extra page will not count against page length
%\bibliographyfullrefs{scibib}

\end{document}